# ABOUT EXACT SOLUTION OF THE KINETIC EQUATION FOR CURRENT SHEATH

V.V. Lyahov[1], V.M. Neshchadim

Institute of Ionosphere, Kamenskoe plato, 050020, Almaty, Kazakhstan

## 1. Introduction

In 1962 Harris offered a distribution function for description tensor of dielectric permeability of the equilibrium current sheath that separates areas with opposite magnetic fields [1]. The distribution function is selected in a manner that allows passing to the system of coordinates where the polarizing electric field is equal to zero. In 1966 Coppi, Laval and Pellat explored kinetic instability of the offered model of equilibrium current sheath [2]. A conclusion was made that under certain conditions tearing instability that courses re-connection magnetic field lines in the magnitoneutral plane may develop. These two papers were basis for all further examinations of dynamics of a current sheath in the geomagnetic tail and energetics of magnetospheric substorms. Equilibrium current sheathes thus are deemed to be electroneutral. Considerations of polarisation of plasma of sharply non-homogeneous current sheaths lie outside of such approach.

A complex of ground and satellite measurements carried out during the recent 15-20 years testifies to inconformity of those theoretical models to real processes that take place in plasma of the near-Earth space. So, the theory cannot explain such a high power of geomagnetic disturbances. Besides, tearing instability in conditions of current sheath in the geomagnetic tail was found to be difficult.

In the offered paper the kinetic theory of equilibrium and stability of the nonelectroneutral current sheaths formed at arbitrary values of medium parameters was developed. Equilibrium distribution function of a current sheath (Harris's function) considering anisotropy of temperature of plasma along and across the layers is summarized. The procedure of examination of a current sheath stability with due regard to plasma polarization effect is offered.

Tensor of dielectric permeability is calculated and dispersion equation for investigation of possible instability modes of non-electroneutral current sheath is derived.

Non-electroneutral equilibrium solution for a current sheath in the magnetospheric tail is derived and its instability is explored.

A possible mechanism of magnetic field annihilation in the magnetospheric tail and acceleration of charged particles in the process of magnetospheric substorm development is considered.

## 2. Statement of the problem

Suppose plasma concentrates in plane z=0 (see Fig. 1), and antiparallel magnetic fields concentrate in the upper and lower half-spaces respectively. In this plane the self-consistent current sheath, which separates areas with opposite magnetic fields, forms.

---

[1] Corresponding author, mail: v_lyahov@rambler.ru

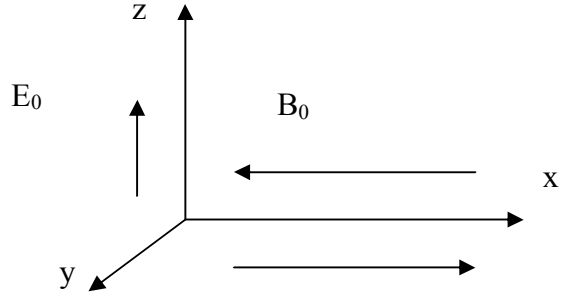

Рис. 1

The problem is one-dimensional. All quantities depend on variable $z$. The medium under study is described by the system of kinetic equation:

$$\frac{\partial f_\alpha}{\partial t} + \vec{v}\frac{\partial f_\alpha}{\partial \vec{r}} + e_\alpha \{\vec{E} + [\vec{v}\vec{B}]\}\frac{\partial f_\alpha}{\partial \vec{P}_\alpha} = 0, \quad (1)$$

and Maxwell equations with self-consistent electromagnetic field (external sourses are absent):

$$rot\vec{B} = \frac{1}{c^2}\frac{\partial \vec{E}}{\partial t} + \frac{1}{\varepsilon_0 c^2}\vec{j}, div\vec{B} = 0, \quad (2)$$

$$rot\vec{E} = -\frac{\partial \vec{B}}{\partial t}, div\vec{E} = \frac{\rho}{\varepsilon_0},$$

where где

$$\rho = \sum_\alpha e_\alpha \int \delta f_\alpha d\vec{P},$$

$$\vec{j} = \sum_\alpha e_\alpha \int \vec{v}\delta f_\alpha d\vec{P}.$$

Problem (1), (2) is solved using the perturbation method

$$f_\alpha(\vec{P},\vec{r},z,t) = f_{0\alpha}(\vec{P}) + \delta f_\alpha(\vec{P},\vec{r},z,t);$$

$$\vec{E}(\vec{r},z,t) = \vec{E}_0(z) + \delta\vec{E}(\vec{r},z,t); \quad (3)$$

$$\vec{B}(\vec{r},z,t) = \vec{B}_0(z) + \delta\vec{B}(\vec{r},z,t).$$

Plasma is deemed to be weakly non-equilibrium $\delta f_\alpha(\vec{P},\vec{r},z,t) < f_{0\alpha}(\vec{P})$.

Equilibrium distribution function is constructed as function of motion integrals $f_{0\alpha}(\vec{P}) = f_{0\alpha}(W_\alpha, P_{y\alpha}, P_x)$ where the total energy and generalized momenta look like:

$$W_\alpha = \frac{1}{2}m_\alpha(v_x^2 + v_y^2 + v_z^2) + e_\alpha\phi(z),$$
$$P_{y\alpha} = m_\alpha v_y + e_\alpha A_y(z), \qquad (4)$$
$$P_x = m_\alpha v_x.$$

Here, $\phi(z), A_y(z)$ are electrical and magnetic potentials ($\vec{E}_0 = -grad\phi, \vec{B}_0 = rot\vec{A}$).

There are some papers, for example [3, 4], which are devoted to the study of polarization of current sheaths. The electric field is calculated in the approximation of quasi-neutrality. The electrical potential is not set equal to zero a priori; it is deduced from the conditions of equality of densities of electrons and ions:

$$n_e(A(z), \phi(z)) = n_i(A(z), \phi(z)). \qquad (5)$$

This relation may be used to express electrical potential through magnetic one $\phi(z) = \phi(A(z))$ and to solve then the remained equation for magnetic potential.

However, requirement (5) is also a restriction imposed on plasma, though it is physically justified. The quasi-neutrality condition is well met at low temperatures and small velocities of macroscopic plasma streams and is worse met at sufficiently elevated temperatures and velocities.

In strict sense, electrically neutral perturbation is zero approximation of method of singular perturbances.

Series expansion parameter for thermal plasma is the ratio of temperature of electrons or ions to electron or ion rest energy, respectively. Obtaining of the next first approximation that is more precise is technically impossible.

This paper is concerned with theoretical research of current sheath and, accordingly, study of the effect of plasma polarization without any restrictions and in a wide range of parameter values. Such statement of the problem goes back to the papers devoted to research of polarization of sharply unhomogeneous magnetoactive plasma [5-8].

3. **Examination of equilibrium of onelectroneutral current sheath**

To describe the equilibrium of current sheath, we will extend Harris's distribution function: we will add anisotropy of temperatures of plasma components along and across the layer, namely:

$$f_{0\alpha} = \frac{m_\alpha}{2\pi\theta_{\alpha z}} n_0 (1+\alpha_\alpha)^{\frac{1}{2}} \exp\left[\frac{m_\alpha}{2\theta_{\alpha z}}(1+\alpha_\alpha)U_\alpha^2\right] \cdot \exp\left[-\frac{W_\alpha}{\theta_{\alpha z}} - \frac{\alpha_\alpha P_{y\alpha}^2}{2\theta_{\alpha z}m_\alpha} + U_\alpha(1+\alpha_\alpha)\frac{P_{y\alpha}}{\theta_{\alpha z}}\right]. \qquad (6)$$

Here,

$$\alpha_\alpha = \frac{\theta_{\alpha z}}{\theta_{\alpha y}} - 1 \text{ - degree of temperature anisotropy,} \qquad (7)$$

$U_\alpha$ – macroscopic velocity along $y$-axis. (8)

As a boundary condition we will take:

$$\phi(z = 0) = A_y(z = 0) = 0. \tag{9}$$

Distribution function (6) on boundary $z = 0$ represents the anisotropic shifted Maxwell distribution:

$$f_{0\alpha} = \frac{m_\alpha}{2\pi\theta_{\alpha z}} n_0 (1+\alpha_\alpha)^{\frac{1}{2}} \exp\left[-\frac{m_\alpha}{2\theta_{\alpha z}}[V_{\alpha z}^2 + (1+\alpha_\alpha)(V_{\alpha y} - U_\alpha)^2]\right]. \tag{10}$$

Having calculated density of particles and current density for equilibrium distribution function (6), it is possible to obtain potential equations:

$$\frac{d^2\phi}{dz^2} = -\frac{en_0}{\varepsilon_0}[\exp(-\frac{e\phi}{\theta_{iz}} + \frac{eU_i A_y}{\theta_{iz}} - \frac{\alpha_i}{(1+\alpha_i)}\frac{e^2 A_y^2}{2\theta_{iz} m_i}) - \\
- \exp(\frac{e\phi}{\theta_{ez}} - \frac{eU_e A_y}{\theta_{ez}} - \frac{\alpha_e}{(1+\alpha_e)}\frac{e^2 A_y^2}{2\theta_{ez} m_e})], \tag{11}$$

$$\frac{d^2 A_y}{dz^2} = -\frac{en_0}{\varepsilon_0 c^2}[(U_i - \frac{\alpha_i}{(1+\alpha_i)}\frac{eA_y}{m_i})\exp(-\frac{e\phi}{\theta_{iz}} + \frac{eU_i A_y}{\theta_{iz}} - \frac{\alpha_i}{(1+\alpha_i)}\frac{e^2 A_y^2}{2\theta_{iz} m_i}) - \\
- (U_e + \frac{\alpha_e}{(1+\alpha_e)}\frac{eA_y}{m_e})\exp(\frac{e\phi}{\theta_{ez}} - \frac{eU_e A_y}{\theta_{ez}} - \frac{\alpha_e}{(1+\alpha_e)}\frac{e^2 A_y^2}{2\theta_{ez} m_e})]. \tag{12}$$

Equations (11), (12) coincide with Harris's equations for the case of isotropic plasma $\alpha_e = \alpha_i = 0$.

$$\frac{d^2\phi}{dz^2} = -\frac{en_0}{\varepsilon_0}[\exp(\frac{eV_i}{\theta_i} A_y - \frac{e}{\theta_i}\phi) - \exp(-\frac{eV_e}{\theta_e} A_y + \frac{e}{\theta_e}\phi)], \tag{13}$$

$$\frac{d^2 A_y}{dz^2} = -\frac{en_0}{\varepsilon_0 c^2}[V_i \exp(\frac{eV_i}{\theta_i} A_y - \frac{e}{\theta_i}\phi) - V_e \exp(-\frac{eV_e}{\theta_e} A_y + \frac{e}{\theta_e}\phi)]. \tag{14}$$

Derivation of an electroneutral solution of the problem for magnetic potential (12) with boundary conditions (9) is possible only by imposing the conditions on the problem parameters:

$$\frac{U_e}{\theta_{ez}} = -\frac{U_i}{\theta_{iz}}, \tag{15}$$

$$\frac{\alpha_e}{(1+\alpha_e)}\frac{1}{\theta_{ez} m_e} = \frac{\alpha_i}{(1+\alpha_i)}\frac{1}{\theta_{iz} m_i}. \tag{16}$$

To derive a neutral solution it is not enough now to pass into the system of coordinates where only condition (15) is fulfilled as it is true of Harris's solution.

The electroneutral solution occurs only for a narrow range of parameter values related to conditions (16).

### 3.1. Numerical investigation of solution for current sheath

*Nonelectroneutral solution.* Let's derive full (nonelectroneutral) solution of combined equations (11), (12). We will add non-dimensional quantities as follows:

$$\mu = \frac{m}{M}, \gamma = \frac{\theta_{ze}}{\theta_{zi}}, \eta = \frac{\theta_{zi}}{Mc^2}, \psi = \frac{e}{\theta_{zi}}\phi, a = \frac{ec}{\theta_{zi}}A_y, \xi^2 = \frac{e^2 n_0}{\theta_{zi}\varepsilon_0}z^2,$$

$$\omega_i = \frac{U_i}{c}, \omega_e = \frac{U_e}{c}. \tag{17}$$

Thus, equations (11), (12) will take on form:

$$\frac{d^2\psi}{d\xi^2} = \exp(\frac{\psi}{\gamma} - \frac{\omega_e}{\gamma}a - \frac{\alpha_e}{1+\alpha_e}\frac{\eta a^2}{2\gamma\mu}) - \exp(-\psi + \omega_i a - \frac{\alpha_i}{1+\alpha_i}\frac{\eta a^2}{2}), \tag{18}$$

$$\frac{d^2 a}{d\xi^2} = (\omega_e + \frac{\alpha_e}{1+\alpha_e}\frac{\eta a}{\mu})\exp(\frac{\psi}{\gamma} - \frac{\omega_e}{\gamma}a - \frac{\alpha_e}{1+\alpha_e}\frac{\eta a^2}{2\gamma\mu}) - (\omega_i - \frac{\alpha_i}{1+\alpha_i}\eta a) \cdot$$

$$\exp(-\psi + \omega_i a - \frac{\alpha_i}{1+\alpha_i}\frac{\eta a^2}{2}). \tag{19}$$

As the boundary conditions we will take:

$$a(\xi = 0) = a'(\xi = 0) = 0;$$
$$\psi(\xi = 0) = \psi'(\xi = 0) = 0. \tag{20}$$

The numerical solution of problem (18), (19), (20) is implemented in package MAPLE 13 by application of the technique of integration of ordinary differential equations *adamsfull* and three integration techniques of stiff equations: *backfull*, *backdiag* and *backfunc*. Integration results completely coincide, and calculation using stiff techniques enables almost double saving of time.

**4. Examination of stability of non-electroneutral current sheath**
**4.1. Determination of nonequilibrium addition to a distribution function $\delta f_\alpha$ and calculation of tensor of dielectric permeability**

Knowing now equilibrium distribution function (5), it is possible to deduce equation for determination of nonequilibrium addition to distribution function $\delta f_\alpha$. Having substituted decompositions (3) in equation (1), and taking into consideration smallness of non-equilibrium(nonequilibrium) components, we will derive a linearized kinetic equation for $\delta f_\alpha$:

$$\frac{\partial \delta f_\alpha}{\partial t} + \vec{v}\frac{\partial \delta f_\alpha}{\partial \vec{r}} + e_\alpha\{\vec{E}_0 + [\vec{v}\vec{B}_0]\}\frac{\partial \delta f_\alpha}{\partial \vec{P}_\alpha} = -e_\alpha\{\delta\vec{E} + [\vec{v}\delta\vec{B}]\}\frac{\partial f_{0\alpha}}{\partial \vec{P}_\alpha}. \tag{21}$$

After decomposition on plane waves the solution of this equation for Fourier's amplitudes will be expression:

$$\delta f_\alpha = \frac{e_\alpha}{\Omega_\alpha(z)} \int_\infty^\varphi (\delta \vec{E} \frac{\partial f_{0\alpha}}{\partial \vec{P}_\alpha})_{\varphi'} \exp[\frac{i}{\Omega_\alpha(z)} \int_\varphi^{\varphi'} (\omega - \vec{k}\vec{v})_{\varphi''} d\varphi''] d\varphi' -$$
$$- \frac{n_0 e_\alpha E_{0z}}{2\pi m_\alpha^2 \theta_\alpha^2 \Omega_\alpha(z)} \exp[-\frac{(P_{\alpha\perp} - \frac{e_\alpha E_{0z} \cos\varphi}{\Omega_\alpha(z)})^2}{2 m_\alpha \theta_\alpha}].$$

(22)

Difference of the obtained solution from the well-known solution for the homogeneous magnetoactive plasma consists, first, in the presence of second term in expression (22). It is caused by the presence of equilibrium electric field. Second, the equilibrium distribution function $f_{0\alpha}$ of the first addend is rigorous self-consistent solution (6) of the stationary kinetic equation. This solution can describe the arbitrary current sheath which character is defined by values of input parameters.

Thus, application of perturbation technique (3) to a kinetic equation with self-consistent field (1), (2) made it possible to find sequentially an equilibrium distribution function (6) and the nonequilibrium addition to it (22).

Theory of current sheath equilibrium is based on equilibrium distribution. The nonequilibrium distribution function (22) may be accepted as a basis for research of current sheath stability.

The discovered nonequilibrium correction to distribution function $\delta f_\alpha$ (22) was used for calculation of the tensor of dielectric permeability (see Application). С помощью найденной неравновесной поправки к функции распределения $\delta f_\alpha$ (22) вычислен тензор диэлектрической проницаемости (см. Приложение).

### 4.2. Procedure of research of dispersion properties of current sheath perturbations

Research of amplitude-frequency characteristics of current sheath perturbations is based on the solution of Maxwell equations for electromagnetic field perturbations closed by constitutive equation:

$$\begin{cases} c^2 \varepsilon_0 rot \delta \vec{B} = \frac{\partial \delta \vec{D}}{\partial t} + \vec{j}_E, div \delta \vec{B} = 0, \\ rot \delta \vec{E} = -\frac{\partial \delta \vec{B}}{\partial t}, div \delta \vec{D} = \rho_E, \end{cases}$$

(23)

where, $\delta \vec{B}(\vec{r},z,t)$ and $\delta \vec{E}(\vec{r},z,t)$ - perturbations of magnetic induction and electric field intensity of the system under consideration; $\delta D(\vec{r},z,t)$ - electric induction perturbation; $\vec{j}_E$ - current density, $\rho_E$ - charge density caused by plasma polarization.

Here, Maxwell equations are closed by constitutive equation:

$$\delta D_i(t,\vec{r},z) = \varepsilon_0 \int_{-\infty}^t dt' \int d\vec{r}' \mathcal{E}_{ij}(t-t',\vec{r}-\vec{r}',z) \delta E_j(t',\vec{r}',z).$$

(24)

By substituted in this equation of plane-wave expansion, we will obtain a constitutive equation for Fourier's amplitudes

$$\delta D_i = \varepsilon_0 \varepsilon_{ij}(\omega, \vec{k}, z) \delta E_j. \tag{25}$$

Here,

$$\varepsilon_{ij}(\omega, \vec{k}, z) = \int_0^\infty dt_1 \int d\vec{r}_1 \varepsilon_{ij}(t_1, \vec{r}_1, z) \exp(i\omega t_1 - i\vec{k}\vec{r}_1) - \tag{26}$$

- tensor of dielectric permeability, $t_1 = t - t'$, $\vec{r}_1 = \vec{r} - \vec{r}'$. We calculated this tensor above (see Application).

Current density $\vec{j}_E$, caused by the presence of equilibrium electric field $E_{0z}$, is determined from the expression:

$$j_{Ey} = -\sum_\alpha \frac{e_\alpha^2 n_0 E_{0z}}{2\pi m_\alpha^2 \theta_\alpha^2 \Omega_\alpha(z)} \int d\vec{P} v_y \exp\left(-\frac{(P_{\alpha\perp} - \frac{e_\alpha E_{0z} \cos\varphi}{\Omega_\alpha(z)})^2}{2 m_\alpha \theta_\alpha}\right) = |v_y = v_\perp \cos\varphi| = \tag{27}$$

$$= -\sum_\alpha \frac{e_\alpha^2 n_0 E_{0z}}{2\pi m_\alpha^2 \theta_\alpha^2 \Omega_\alpha(z)} \int_0^\infty dP_{\alpha\perp} v_\perp P_{\alpha\perp} \int_0^{2\pi} d\varphi \cos\varphi \exp\left(-\frac{(P_{\alpha\perp} - \frac{e_\alpha E_{0z} \cos\varphi}{\Omega_\alpha(z)})^2}{2 m_\alpha \theta_\alpha}\right) \int_{-\infty}^\infty dP_{\alpha x}.$$

Integration of the last expression for the nonrelativistic plasma yields the result:

$$j_{Ey}(z) = -\sum_\alpha \frac{\Gamma(\frac{1}{4}) e_\alpha^{\frac{7}{2}} n_0 c}{\sqrt{2\pi} \theta_\alpha^{\frac{5}{4}} m_\alpha^{\frac{5}{4}}} \left(\frac{E_{0z}(z)}{\Omega_\alpha(z)}\right)^{\frac{5}{2}}, \tag{28}$$

$$j_{Ex} = j_{Ez} = 0. \tag{29}$$

We solve initial problem. In spatially inhomogeneous unbounded medium, by the initial instant $t=0$ stationary electromagnetic field $E_{0z}(z)$, $B_{0z}(z)$ and relative stationary current density $\vec{j}_E(z)$ and charge density $\rho_E(z)$ were created. After substitution of Fourier transforms:

$$\vec{E}(t, \vec{r}) = \int_{-\infty+i\delta}^{\infty+i\delta} d\omega \exp(-i\omega t) \int_{-\infty}^{+\infty} d\vec{k} \exp(i\vec{k}\vec{r}) \vec{E}(\omega, \vec{k}),$$

$$\vec{B}(t, \vec{r}) = \int_{-\infty+i\delta}^{\infty+i\delta} d\omega \exp(-i\omega t) \int_{-\infty}^{+\infty} d\vec{k} \exp(i\vec{k}\vec{r}) \vec{B}(\omega, \vec{k}),$$

$$\vec{D}(t, \vec{r}) = \int_{-\infty+i\delta}^{\infty+i\delta} d\omega \exp(-i\omega t) \int_{-\infty}^{+\infty} d\vec{k} \exp(i\vec{k}\vec{r}) \vec{D}(\omega, \vec{k}),$$

$$\vec{j}_E(t=0, \vec{r}) = \int_{-\infty+i\delta}^{\infty+i\delta} d\omega \exp(-i\omega \cdot (t=0)) \int_{-\infty}^{+\infty} d\vec{k} \exp(i\vec{k}\vec{r}) \vec{j}_E(t=0, \vec{k}), \tag{30}$$

$$\rho_E(t=0,\vec{r}) = \int\limits_{-\infty+i\delta}^{\infty+i\delta} d\omega \exp(-i\omega \cdot (t=0)) \int\limits_{-\infty}^{+\infty} d\vec{k} \exp(i\vec{k}\vec{r}) \rho_E(t=0,\vec{k}).$$

we obtain from Maxwell equations (23) an equation for electric field perturbation

$$[k^2 \delta_{ij} - k_i k_j - \frac{\omega^2}{c^2} \varepsilon_{ij}(\omega,\vec{k})] \delta E_j = \frac{i\omega}{\varepsilon_0 c^2} j_{Ei}(t=0,\vec{k}), \qquad (31)$$

where tensor of dielectric permeability $\varepsilon_{ij}(\omega,\vec{k})$ is defined in Application, and the Fourier transform on spatial coordinates from initial value of current density looks like:

$$j_{Ei}(t=0,\vec{k}) = \frac{1}{(2\pi)^3} \int\limits_{-\infty}^{+\infty} d\vec{r} \exp(-i\vec{k}\vec{r}) j_{Ei}(t=0,\vec{r}). \qquad (32)$$

The initial current density $j_{Ei}(t=0,\vec{r})$ depends only on coordinate *z* and its components are defined by expressions (28), (29).

The equation (31) represents system of three inhomogeneous algebraic equations. Examination of dispersion properties of perturbations of a boundary layer is based on the solution of inhomogeneous system (31). Method of solution of those equations consists in the following [9], p. 149.

Let's write the solutions using Cramer formulas:

$$\delta E_x = \frac{D_x}{D}, \quad \delta E_y = \frac{D_y}{D}, \quad \delta E_z = \frac{D_z}{D}. \qquad (33)$$

Here $D = [k^2 \delta_{ij} - k_i k_j - \frac{\omega^2}{c^2} \varepsilon_{ij}(\omega,\vec{k})]$ - \qquad (34)

- determinant of system (31), $D_i$ – the determinant obtained by replacement in determinant (34) of i-th column by a column from absolute terms of the equation (31).

Further it is necessary to resort to numerical calculation. Among solutions of system (31) there may be such solutions which all four determinants simultaneously turn into zero.

$$D = 0, \quad D_x = 0, \quad D_y = 0, \quad D_z = 0. \qquad (35)$$

If the numerical solution of system (35) gives at least one dependence $\omega = f_1(\vec{k})$, it would be necessary to check this solution. In this case rank of the matrix

$$\begin{Vmatrix} a_{11}, a_{12}, a_{13} \\ a_{21}, a_{22}, a_{23} \\ a_{31}, a_{32}, a_{33} \end{Vmatrix} \qquad (36)$$

should be equal to the rank of extended matrix

$$\begin{Vmatrix} a_{11}, a_{12}, a_{13}, b_1 \\ a_{21}, a_{22}, a_{23}, b_2 \\ a_{31}, a_{32}, a_{33}, b_3 \end{Vmatrix}. \tag{37}$$

Here, $a_{ij}$ - coefficients of the equation (31) which become defined after substitution of the found solution $\omega = f_1(\vec{k})$, $b_i$ - components of the inhomogeneous side of equation (31) also defined after substitution of the solution. So, the solution of system (35) defines dispersion characteristics of current sheath perturbations.

Subintegral functions of permittivity tensor components are rapidly oscillating functions that represent major complexity in numerical calculation. MAPLE-13 Package is used for solution of dispersion equation (31) based on Monte-Carlo method. It is known that the Monte-Carlo method is a slowly converging procedure. Investigation turned out that results of evaluation vary a little at transition of random number generation from j = 1000 to j = 10000.

Thus, for j = 1000 relative error of the calculations which have been evaluated based on procedure [10], p. 238, was $\delta = 3\%$.

## 5. Application of the theory to investigation of equilibrium and stability of current sheath of magnetospheric tail

Let's apply the procedure of investigation of equilibrium of the non-electroneutral current sheaths developed in Section 3 to current sheath of the magnetospheric tail. System of equations for electromagnetic potentials (18), (19) with boundary conditions (20) is numerically solved at parameter values that are typical of the area of magnetospheric tail:

$$\mu = 0.00055, \eta = 1.07 \cdot 10^{-5}, \gamma = 0.1. \tag{38}$$

Remaining 4 arbitrary parametres are selected close to their electroneutral values defined from requirements (15, 16):

$$\omega_e = -\gamma \omega_i, \tag{39}$$

$$\frac{\alpha_e}{1+\alpha_e} = \mu\gamma \frac{\alpha_i}{1+\alpha_i}, \tag{40}$$

namely:

$$\omega_i = 0.33 \cdot 10^{-2}, \omega_e = -0.33 \cdot 10^{-3}, \alpha_i = 1.0 \cdot 10^{-3}, \alpha_e = 5.5 \cdot 10^{-7}. \tag{41}$$

If conditions (39), (40) are met, the solution of combined equations (18), (19), (20) leads to the known Harris's electroneutral equilibrium which is only a special case of more common nonelectroneutral solution. Space plasma possesses arbitrary values of the parameters which generally do not obey to requirements (39), (40). In this case it appears that all solutions, at first, are nonelectroneutral. Fig. 2a shows a magnetic potential profile; Fig. 2b – electrical profile; Fig. 2c - magnetic field profile; Fig. 2d – electrical profile. Fig. 2e shows the balance of pressures where function:

$$R(z) = \frac{\varepsilon_0 E_0^2(z)}{2} + \frac{B_0^2(z)}{2\mu_0} + n_0(z) \cdot (\theta_i + \theta_e) - (\theta_i + \theta_e) \cdot n_0(z=0), \tag{42}$$

representing deviation of the pressure at arbitrary point z from the pressure in magnitoneutral plane z=0 is plotted on the ordinate.

Second, at some distance from the magnitoneutral plane $z = 0$ the electrical potential and polarising electric field sharply increase in some very narrow area. Fig. 3 shows such solutions. Curves in Fig. 3 differ from the curves in Fig. 2 by value of only one parameter; for them: $\alpha_e = 5.5 \cdot 10^{-6.7}$. All solutions in Figs. 2 and 3 are obtained for the characteristic depth:

$$z = \sqrt{\frac{\theta_{zi}\varepsilon_0}{e^2 n_0}} = 1859 \text{ m.} \tag{43}$$

The balance of pressures fulfilled with good precision at all points of the nonelectroneutral current sheath is sharply broken in this narrow area owing to electric field growth. In strict sense the obtained solution is not equilibrium, but it is possible to consider it as a certain "quasi-equilibrium" solution which is obviously unstable.

The procedure of investigation of instability of the nonelectroneutral current sheaths developed in section 4 allows to obtain all instability modes of the current sheath under consideration. Thus, at values of parameters (38), (41) and wave vectors $k_x = k_y = =0.0005$ each of four equations (35) gives accordingly 108, 63, 86 and 84 roots. Of which solutions with positive real parts have physical sense only if those solutions satisfy all equations (35) at a time and the rank of basic matrix (36) is equal to the rank of extended matrix (37). One of such solutions is:

$$\omega = 45 + 4i. \tag{44}$$

It is natural that the most destructive are modes with the greatest positive imaginary parts.

### 5.1. About magnetic field annihilation in the current sheath of magnetospheric tail and acceleration of the charged particles in the process of magnetospheric substorm

In Section 3 it was noted that the complete nonelectroneutral equilibrium solution is structurally unstable and can qualitatively vary at small change of the values of parameters. At the same time Harris' solution is structurally stable. This difference leads to the fact that owing to fluctuation of medium parameters the nonelectroneutral equilibrium can occur only in the space volumes that are much less than areas where Harris' electroneutral equilibrium appears.

Thus, the effect of polarization of magnetoactive plasma in the magnetospheric tail leads to dissection of all the space of central, core area of the tail into separate domains, in each of which at a certain time its own sets of medium parameters appear: $\alpha_e, \alpha_i, \gamma, \eta, \omega_e, \omega_i, \mu$. Depending on the values of those parameters, in each domain its own quasiequilibrium with various equilibrium distances (Fig. 2, Fig. 3) appears. Thus, the further values of parameters decline from the electroneutral values (39), (40), the narrower the equilibrium zone is. The excess of pressure on the periphery of a quasi-equilibrium current sheath in each domain leads to quick collapse of the current sheath structure to the center, i.e. to the magnitoneutral plane and annihilation of the magnetic field. Intensity of the process is determined by the increment of most destructive mode of current sheath instability in this domain. In various domains the annihilation has different intensity. All domains rearrange with time, i.e. destroy, merge with other domains, etc. During the development of magnetospheric substorm, density of current sheath plasma and magnetic field in the magnetospheric tail increase and it leads to increase of intensity of magnetic field annihilation.

Charged particles of plasma in the vicinity of magnitoneutral plane appear to be captured in the magnetic trap (see Fig. 3). Those particles are accelerated by means of Fermi's mechanism during rapid contraction of the strong peripheral magnetic field to the center. Kinetic energy caused by transverse (with reference to the current sheath) velocity of particles grows. Allowing for real inclination of magnetic field lines to magnitoneutral plane, going out of those accelerated particles along magnetic field lines to the area of polar oval of the Earth is possible. The offered mechanism of acceleration needs to be quantitatively analyzed, but it is quite possible that such mechanism leads to more effective acceleration of particles during magnetospheric substorms than mechanism discussed earlier and related to acceleration of particles during their oscillations along the magnetic field line between the advancing points of reflections of a particle in the core of dipole and the magnitoneutral plane of magnetospheric tail.

**6. Conclusion**

For description of a current sheath the preference should be given to equilibrium distribution function (6) in comparison to distribution function of Harris widely used now because allocation (6) allows description of more common situation when temperatures of components of current sheath plasma are anisotropic. Equilibrium of current sheath when plasma parameters do not fulfill certain requirements (15), (16) is nonelectroneutral.

The procedure of investigation of instability of a nonelectroneutral current sheath has been created. On the basis of the obtained nonequilibrium addition to the distribution function the permittivity tensor simulating the medium (current sheath) was calculated. This tensor essentially differs from the known tensor for the homogeneous magnetoactive plasma.

Application of the nonelectroneutral theory to investigation of the equilibrium and stability of the magnetospheric tail leads to an alternative pattern (in comparison with the electroneutral model based on Harris' stationary distribution). First, the equilibrium current sheath has domain structure; each domain is defined by its own set of parameters. Second, all domains possess a polarization electric field that sharply increases on the periphery (far from the magnitoneutral plane) forming some quasi-equilibrium structure. Third, rate of annihilation of a magnetic field is highe, than in the electroneutral model and is defined by the velocity of collapse to the center of domain structure. Fourth, rate of acceleration of the charged particles is higher than in the standard electroneutral model. Nonelectroneutral (created taking into account the effect of plasma polarisation) theory seems to be more adequate to the accumulated complex of ground based and satellite measurements.

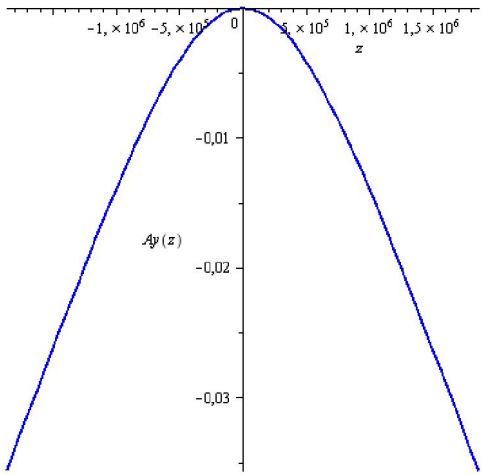

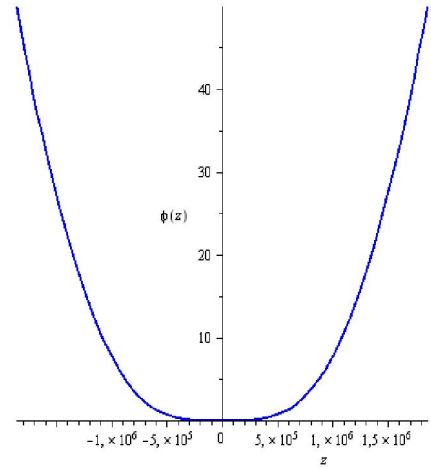

Fig. 2a

Fig. 2b

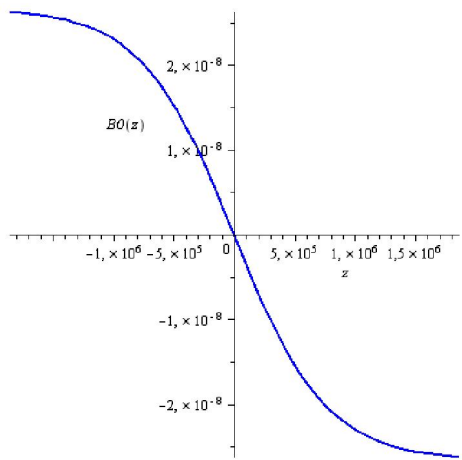

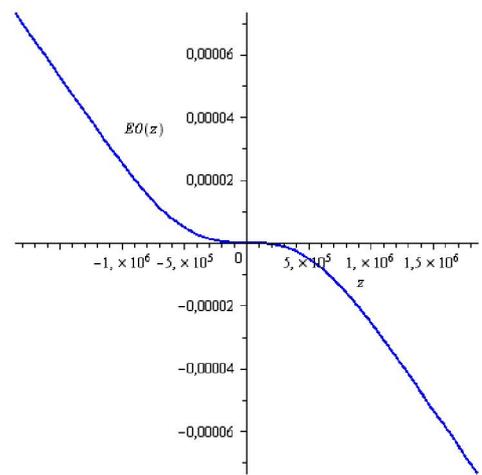

Fig. 2c

Fig. 2d

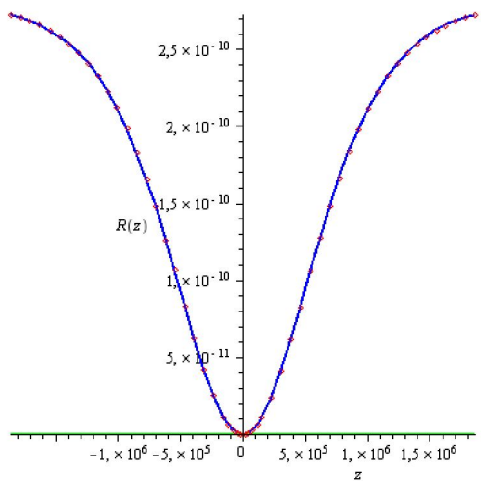

Fig.. 2e

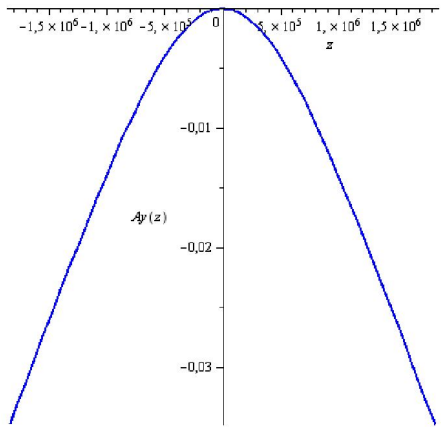

Fig.3a

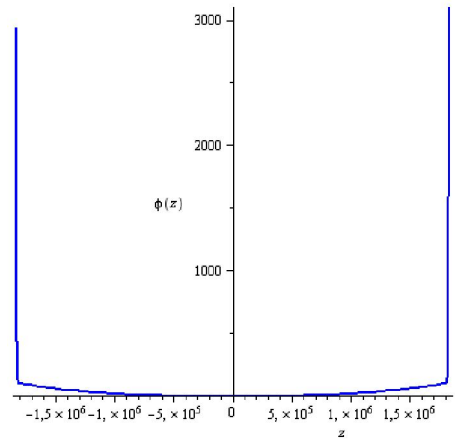

Fig. 3b

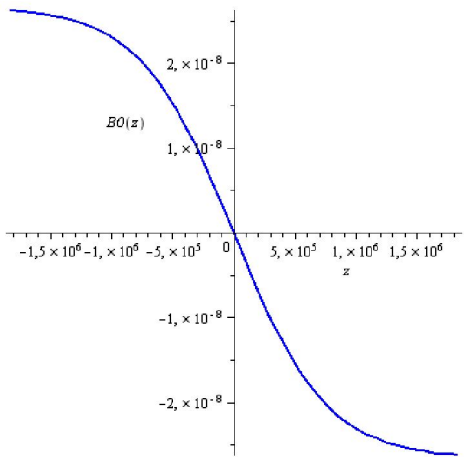

Fig. 3c

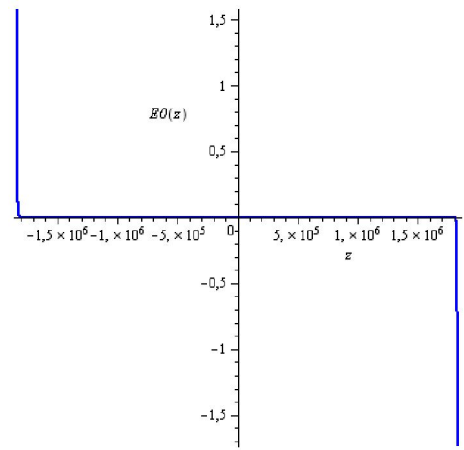

Fig. 3d

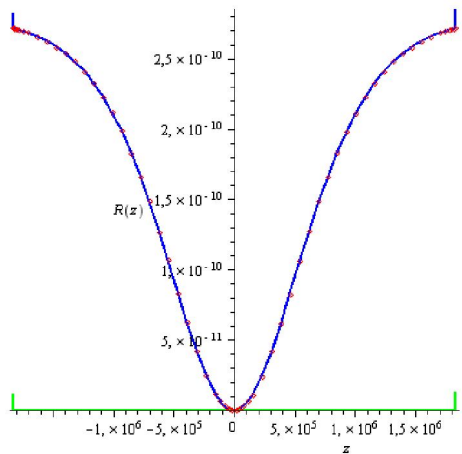

Fig. 3e

# APPLICATION

**Anywhere:**

$$B_\alpha = \frac{2\pi e_\alpha^2}{\varepsilon_0 \omega} \left(\frac{m_\alpha}{2\pi\theta_\alpha}\right)^{\frac{3}{2}} \frac{n_{0\alpha}}{\theta_\alpha} (1+\alpha_\alpha)^{\frac{1}{2}} \exp\left(-\frac{m_\alpha}{2\theta_\alpha}(1+\alpha_\alpha)U_\alpha^2\right) \cdot$$
$$\exp\left[-\frac{m_\alpha}{2\theta_\alpha}\left(\frac{2e_\alpha\phi(z)}{m_\alpha} + \frac{\alpha_\alpha e_\alpha^2 A_y^2(z)}{m_\alpha^2} - \frac{2U_\alpha(1+\alpha_\alpha)e_\alpha A_y(z)}{m_\alpha}\right)\right].$$

(A1)

## TENSOR COMPONENTS

$$\varepsilon_{xx} = 1 - \sum_\alpha B_\alpha \sum_s \sum_l G_{xx}(s,l,m_\alpha,\theta_\alpha,k_\perp,\Omega_\alpha(z)) I_{xx}(\beta). \tag{A2}$$

Here

$$G_{xx}(s,l,m_\alpha,\theta_\alpha,k_\perp,\Omega_\alpha(z)) = \int_0^\infty C_s^{xx} v_\perp \exp\left(-\frac{m_\alpha v_\perp^2}{2\theta_\alpha}\right) J_{s+l}\left(\frac{k_\perp v_\perp}{\Omega_\alpha(z)}\right) J_l\left(\frac{k_\perp v_\perp}{\Omega_\alpha(z)}\right) dv_\perp$$

$$I_{xx}(\beta) = \int_{-\infty}^{+\infty} \frac{v_x^2 \exp\left(-\frac{m_\alpha v_x^2}{2\theta_\alpha}\right)}{(\omega - k_x v_x - l\Omega_\alpha(z))} dv_x = \frac{\theta_\alpha}{m_\alpha k_x}\left[\frac{\sqrt{2\pi}}{\beta} - i\pi\beta^2 \exp\left(-\frac{\beta^2}{2}\right)\right], \text{если } \beta \gg 1$$

$$C_s^{xx} = \frac{2}{\pi} \int_{\frac{\pi}{2}(\alpha-1)}^{\frac{\pi}{2}\alpha} \exp(-i4s\varphi') \exp\left\{-\frac{m_\alpha}{2\theta_\alpha}\left[\alpha_\alpha v_\perp^2 \cos^2\varphi' + 2\left(\alpha_\alpha \frac{e_\alpha}{m_\alpha} A_y(z) - U_\alpha(1+\alpha_\alpha)\right)v_\perp \cos\varphi'\right]\right\} d\varphi'.$$

$$\varepsilon_{xy} = -\sum_\alpha B_\alpha \sum_s \sum_l G_{xy}(s,l,m_\alpha,\theta_\alpha,k_\perp,\Omega_\alpha(z)) I_{xy}(\beta). \tag{A3}$$

Here

$$G_{xy}(s,l,m_\alpha,\theta_\alpha,k_\perp,\Omega_\alpha(z)) = \int_0^\infty C_s^{xy} v_\perp \exp\left(-\frac{m_\alpha v_\perp^2}{2\theta_\alpha}\right) J_{s+l}\left(\frac{k_\perp v_\perp}{\Omega_\alpha(z)}\right) J_l\left(\frac{k_\perp v_\perp}{\Omega_\alpha(z)}\right) dv_\perp,$$

$$I_{xy}(\beta) = \int_{-\infty}^{+\infty} \frac{v_x \exp\left(-\frac{m_\alpha v_x^2}{2\theta_\alpha}\right)}{(\omega - k_x v_x - l\Omega_\alpha(z))} dv_x \approx \frac{1}{k_x}\sqrt{\frac{\theta_\alpha}{m_\alpha}} \frac{\sqrt{2\pi}}{\beta^2}, \beta \gg 1,$$

$$C_s^{xy} = \frac{2}{\pi} \int_{\frac{\pi}{2}(\alpha-1)}^{\frac{\pi}{2}\alpha} \exp(-i4s\varphi') \cdot [(1+\alpha_\alpha)v_\perp \cos\varphi' + \frac{\alpha_\alpha e_\alpha}{m_\alpha} A_y(z) - U_\alpha(1+\alpha_\alpha)] \cdot$$

$$\exp\{-\frac{m_\alpha}{2\theta_\alpha}[\alpha_\alpha v_\perp^2 \cos^2\varphi' + 2(\alpha_\alpha \frac{e_\alpha}{m_\alpha} A_y(z) - U_\alpha(1+\alpha_\alpha))v_\perp \cos\varphi']\}d\varphi'.$$

$$\varepsilon_{xz} = -\sum_\alpha B_\alpha \sum_s \sum_l G_{xz}(s,l,m_\alpha,\theta_\alpha,k_\perp,\Omega_\alpha(z))I_{xz}(\beta). \tag{A4}$$

Here

$$G_{xz}(s,l,m_\alpha,\theta_\alpha,k_\perp,\Omega_\alpha(z)) = \int_0^\infty C_s^{xz} v_\perp^2 \exp(-\frac{m_\alpha v_\perp^2}{2\theta_\alpha}) J_{s+l}(\frac{k_\perp v_\perp}{\Omega_\alpha(z)}) J_l(\frac{k_\perp v_\perp}{\Omega_\alpha(z)}) dv_\perp,$$

$$I_{xz}(\beta) = \int_{-\infty}^{+\infty} \frac{v_x \exp(-\frac{m_\alpha v_x^2}{2\theta_\alpha})}{(\omega - k_x v_x - l\Omega_\alpha(z))} dv_x \approx \frac{1}{k_x}\sqrt{\frac{\theta_\alpha}{m_\alpha}}\frac{\sqrt{2\pi}}{\beta^2}, \quad \beta \gg 1,$$

$$C_s^{xz} = \frac{2}{\pi} \int_{\frac{\pi}{2}(\alpha-1)}^{\frac{\pi}{2}\alpha} \exp(-i4s\varphi') \cdot \sin\varphi' \cdot$$

$$\exp\{-\frac{m_\alpha}{2\theta_\alpha}[\alpha_\alpha v_\perp^2 \cos^2\varphi' + 2(\alpha_\alpha \frac{e_\alpha}{m_\alpha} A_y(z) - U_\alpha(1+\alpha_\alpha))v_\perp \cos\varphi']\}d\varphi'.$$

$$\varepsilon_{yx} = -\sum_\alpha B_\alpha \sum_s \sum_l G_{yx}(s,l,m_\alpha,\theta_\alpha,k_\perp,\Omega_\alpha(z))I_{yx}(\beta). \tag{A5}$$

Here

$$G_{yx}(s,l,m_\alpha,\theta_\alpha,k_\perp,\Omega_\alpha(z)) = \int_0^\infty C_s^{yx} v_\perp^2 \exp(-\frac{m_\alpha v_\perp^2}{2\theta_\alpha}) J_{s+l}(\frac{k_\perp v_\perp}{\Omega_\alpha(z)}) lJ_l(\frac{k_\perp v_\perp}{\Omega_\alpha(z)}) \frac{\Omega_\alpha(z)}{k_\perp v_\perp} dv_\perp,$$

$$I_{yx}(\beta) = \int_{-\infty}^{+\infty} \frac{v_x \exp(-\frac{m_\alpha v_x^2}{2\theta_\alpha})}{(\omega - k_x v_x - l\Omega_\alpha(z))} dv_x \approx \frac{1}{k_x}\sqrt{\frac{\theta_\alpha}{m_\alpha}}\frac{\sqrt{2\pi}}{\beta^2}, \quad \beta \gg 1,$$

$$C_s^{yx} = \frac{2}{\pi} \int_{\frac{\pi}{2}(\alpha-1)}^{\frac{\pi}{2}\alpha} \exp(-i4s\varphi') \cdot$$

$$\exp\{-\frac{m_\alpha}{2\theta_\alpha}[\alpha_\alpha v_\perp^2 \cos^2\varphi' + 2(\alpha_\alpha \frac{e_\alpha}{m_\alpha}A_y(z) - U_\alpha(1+\alpha_\alpha))v_\perp \cos\varphi']\}d\varphi'.$$

$$\varepsilon_{yy} = 1 - \sum_\alpha B_\alpha \sum_s \sum_l G_{yy}(s,l,m_\alpha,\theta_\alpha,k_\perp,\Omega_\alpha(z))I_{yy}(\beta). \tag{A6}$$

Here

$$G_{yy}(s,l,m_\alpha,\theta_\alpha,k_\perp,\Omega_\alpha(z)) = \int_0^\infty C_s^{yy} v_\perp^2 \exp(-\frac{m_\alpha v_\perp^2}{2\theta_\alpha}) J_{s+l}(\frac{k_\perp v_\perp}{\Omega_\alpha(z)}) l J_l(\frac{k_\perp v_\perp}{\Omega_\alpha(z)}) \frac{\Omega_\alpha(z)}{k_\perp v_\perp} dv_\perp,$$

$$I_{yy}(\beta) = \int_{-\infty}^{+\infty} \frac{\exp(-\frac{m_\alpha v_x^2}{2\theta_\alpha})}{(\omega - k_x v_x - l\Omega_\alpha(z))} dv_x \approx \frac{1}{k_x}\frac{\sqrt{2\pi}}{\beta}, \quad \beta \gg 1,$$

$$C_s^{yy} = \frac{2}{\pi} \int_{\frac{\pi}{2}(\alpha-1)}^{\frac{\pi}{2}\alpha} \exp(-i4s\varphi') \cdot [(1+\alpha_\alpha)v_\perp \cos\varphi' + \frac{\alpha_\alpha e_\alpha}{m_\alpha}A_y(z) - U_\alpha(1+\alpha_\alpha)] \cdot$$

$$\exp\{-\frac{m_\alpha}{2\theta_\alpha}[\alpha_\alpha v_\perp^2 \cos^2\varphi' + 2(\alpha_\alpha \frac{e_\alpha}{m_\alpha}A_y(z) - U_\alpha(1+\alpha_\alpha))v_\perp \cos\varphi']\}d\varphi'.$$

$$\varepsilon_{yz} = -\sum_\alpha B_\alpha \sum_s \sum_l G_{yz}(s,l,m_\alpha,\theta_\alpha,k_\perp,\Omega_\alpha(z)) \cdot I_{yz}(\beta). \tag{A7}$$

Here

$$G_{yz}(s,l,m_\alpha,\theta_\alpha,k_\perp,\Omega_\alpha(z)) = \int_0^\infty C_s^{yz} v_\perp^3 \exp(-\frac{m_\alpha v_\perp^2}{2\theta_\alpha}) J_{s+l}(\frac{k_\perp v_\perp}{\Omega_\alpha(z)}) l J_l(\frac{k_\perp v_\perp}{\Omega_\alpha(z)}) \frac{\Omega_\alpha(z)}{k_\perp v_\perp} dv_\perp,$$

$$I_{yz}(\beta) = \int_{-\infty}^{+\infty} \frac{\exp(-\frac{m_\alpha v_x^2}{2\theta_\alpha})}{(\omega - k_x v_x - l\Omega_\alpha(z))} dv_x \approx \frac{1}{k_x}\frac{\sqrt{2\pi}}{\beta}, \quad \beta \gg 1,$$

$$C_s^{yz} = \frac{2}{\pi} \int_{\frac{\pi}{2}(\alpha-1)}^{\frac{\pi}{2}\alpha} \exp(-i4s\varphi') \cdot \sin\varphi' \cdot$$

$$\exp\{-\frac{m_\alpha}{2\theta_\alpha}[\alpha_\alpha v_\perp^2 \cos^2\varphi' + 2(\alpha_\alpha \frac{e_\alpha}{m_\alpha} A_y(z) - U_\alpha(1+\alpha_\alpha))v_\perp \cos\varphi']\}d\varphi'.$$

$$\varepsilon_{zx} = i\sum_\alpha B_\alpha \sum_s \sum_l G_{zx}(s,l,m_\alpha,\theta_\alpha,k_\perp,\Omega_\alpha(z)) \cdot I_{zx}(\beta). \tag{A8}$$

Here

$$G_{zx}(s,l,m_\alpha,\theta_\alpha,k_\perp,\Omega_\alpha(z)) = \int_0^\infty C_s^{zx} v_\perp^2 \exp(-\frac{m_\alpha v_\perp^2}{2\theta_\alpha}) J_{s+l}(\frac{k_\perp v_\perp}{\Omega_\alpha(z)}) J_l'(\frac{k_\perp v_\perp}{\Omega_\alpha(z)}) dv_\perp,$$

where:

$$J_l'(\frac{k_\perp v_\perp}{\Omega_\alpha(z)}) = \frac{1}{2}(J_{l-1}(\frac{k_\perp v_\perp}{\Omega_\alpha(z)}) - J_{l+1}(\frac{k_\perp v_\perp}{\Omega_\alpha(z)})).$$

$$I_{zx}(\beta) = \int_{-\infty}^{+\infty} \frac{v_x \exp(-\frac{m_\alpha v_x^2}{2\theta_\alpha})}{(\omega - k_x v_x - l\Omega_\alpha(z))} dv_x \approx \frac{1}{k_x}\sqrt{\frac{\theta_\alpha}{m_\alpha}}\frac{\sqrt{2\pi}}{\beta^2}, \quad \beta >> 1,$$

$$C_s^{zx} = \frac{2}{\pi} \int_{\frac{\pi}{2}(\alpha-1)}^{\frac{\pi}{2}\alpha} \exp(-i4s\varphi') \cdot$$

$$\exp\{-\frac{m_\alpha}{2\theta_\alpha}[\alpha_\alpha v_\perp^2 \cos^2\varphi' + 2(\alpha_\alpha \frac{e_\alpha}{m_\alpha} A_y(z) - U_\alpha(1+\alpha_\alpha))v_\perp \cos\varphi']\}d\varphi'.$$

$$\varepsilon_{zy} = i\sum_\alpha B_\alpha \sum_s \sum_l G_{zy}(s,l,m_\alpha,\theta_\alpha,k_\perp,\Omega_\alpha(z)) \cdot I_{zy}(\beta). \tag{A9}$$

Here

$$G_{zy}(s,l,m_\alpha,\theta_\alpha,k_\perp,\Omega_\alpha(z)) = \int_0^\infty C_s^{zy} v_\perp^2 \exp(-\frac{m_\alpha v_\perp^2}{2\theta_\alpha}) J_{s+l}(\frac{k_\perp v_\perp}{\Omega_\alpha(z)}) J_l'(\frac{k_\perp v_\perp}{\Omega_\alpha(z)}) dv_\perp,$$

where:

$$J'_l(\frac{k_\perp v_\perp}{\Omega_\alpha(z)}) = \frac{1}{2}(J_{l-1}(\frac{k_\perp v_\perp}{\Omega_\alpha(z)}) - J_{l+1}(\frac{k_\perp v_\perp}{\Omega_\alpha(z)})).$$

$$I_{zy}(\beta) = \int_{-\infty}^{+\infty} \frac{\exp(-\frac{m_\alpha v_x^2}{2\theta_\alpha})}{(\omega - k_x v_x - l\Omega_\alpha(z))} dv_x \approx \frac{1}{k_x}\frac{\sqrt{2\pi}}{\beta}, \quad \beta \gg 1,$$

$$C_s^{zy} = \frac{2}{\pi}\int_{\frac{\pi}{2}(\alpha-1)}^{\frac{\pi}{2}\alpha} \exp(-i4s\varphi') \cdot [(1+\alpha_\alpha)v_\perp \cos\varphi' + \frac{\alpha_\alpha e_\alpha}{m_\alpha}A_y(z) - U_\alpha(1+\alpha_\alpha)] \cdot$$

$$\exp\{-\frac{m_\alpha}{2\theta_\alpha}[\alpha_\alpha v_\perp^2 \cos^2\varphi' + 2(\alpha_\alpha \frac{e_\alpha}{m_\alpha}A_y(z) - U_\alpha(1+\alpha_\alpha))v_\perp \cos\varphi']\}d\varphi'.$$

$$\varepsilon_{zz} = 1 + i\sum_\alpha B_\alpha \sum_s \sum_l G_{zz}(s,l,m_\alpha,\theta_\alpha,k_\perp,\Omega_\alpha(z)) \cdot I_{zz}(\beta). \tag{A10}$$

Here

$$G_{zz}(s,l,m_\alpha,\theta_\alpha,k_\perp,\Omega_\alpha(z)) = \int_0^\infty C_s^{zz} v_\perp^3 \exp(-\frac{m_\alpha v_\perp^2}{2\theta_\alpha}) J_{s+l}(\frac{k_\perp v_\perp}{\Omega_\alpha(z)}) J'_l(\frac{k_\perp v_\perp}{\Omega_\alpha(z)}) dv_\perp,$$

where:

$$J'_l(\frac{k_\perp v_\perp}{\Omega_\alpha(z)}) = \frac{1}{2}(J_{l-1}(\frac{k_\perp v_\perp}{\Omega_\alpha(z)}) - J_{l+1}(\frac{k_\perp v_\perp}{\Omega_\alpha(z)})).$$

$$I_{zz}(\beta) = \int_{-\infty}^{+\infty} \frac{\exp(-\frac{m_\alpha v_x^2}{2\theta_\alpha})}{(\omega - k_x v_x - l\Omega_\alpha(z))} dv_x \approx \frac{1}{k_x}\frac{\sqrt{2\pi}}{\beta}, \quad \beta \gg 1,$$

$$C_s^{zz} = \frac{2}{\pi}\int_{\frac{\pi}{2}(\alpha-1)}^{\frac{\pi}{2}\alpha} \exp(-i4s\varphi') \cdot \sin\varphi' \cdot$$

$$\exp\{-\frac{m_\alpha}{2\theta_\alpha}[\alpha_\alpha v_\perp^2 \cos^2\varphi' + 2(\alpha_\alpha \frac{e_\alpha}{m_\alpha}A_y(z) - U_\alpha(1+\alpha_\alpha))v_\perp \cos\varphi']\}d\varphi'.$$